\documentstyle[epsf,multicol,aps]{revtex}

\begin{document}
\draft
\title{Crossover and self-averaging in the two-dimensional 
site-diluted Ising model:  
Application of probability-changing cluster algorithm}

\author{Yusuke Tomita\cite{tomita} and Yutaka Okabe\cite{okabe}}

\address{
Department of Physics, Tokyo Metropolitan University,
Hachioji, Tokyo 192-0397, Japan
}

\date{Received \today}

\maketitle

\begin{abstract}
Using the newly proposed probability-changing cluster (PCC) 
Monte Carlo algorithm, we simulate the two-dimensional (2D) 
site-diluted Ising model.  Since we can tune the critical point 
of each random sample automatically with the PCC algorithm, 
we succeed in studying the sample-dependent $T_c(L)$ and 
the sample average of physical quantities at each $T_c(L)$ 
systematically. 
Using the finite-size scaling (FSS) analysis for $T_c(L)$, 
we discuss the importance of corrections to FSS 
both in the strong-dilution and weak-dilution regions. 
The critical phenomena of the 2D site-diluted Ising model are 
shown to be controlled by the pure fixed point.  
The crossover from the percolation fixed point 
to the pure Ising fixed point with the system size is 
explicitly demonstrated by the study of the Binder parameter. 
We also study the distribution of critical temperature $T_c(L)$. 
Its variance shows the power-law $L$ dependence, $L^{-n}$, 
and the estimate of the exponent $n$ is consistent with the prediction 
of Aharony and Harris [Phys. Rev. Lett. {\bf 77}, 3700 (1996)].
Calculating the relative variance of critical magnetization 
at the sample-dependent $T_c(L)$, we show that the 2D site-diluted 
Ising model exhibits weak self-averaging. 

\end{abstract}

\pacs{PACS numbers: 75.50.Lk, 64.60.Ak, 05.10.Ln}

\begin{multicols}{2}
\narrowtext

\section{Introduction}

The critical behavior of random spin systems have been studied 
for more than decades \cite{Stinchcombe}.  The effect of randomness 
has been attracting much attention because real materials 
contain impurities.  In a pioneering work, Harris \cite{Harris} 
studied the problem whether the randomness changes the critical 
behavior of the pure system in the case of quenched random systems.  
The so-called Harris criterion \cite{Harris} states that 
even a weak randomness changes the critical behavior 
if the specific heat exponent $\alpha_{\rm pure}$ of the pure system 
is positive; the random system belongs to the different 
universality class from that of the pure system.  
Then, the randomness becomes relevant 
in the renormalization group (RG) terminology; 
the critical behavior is controlled by the random fixed point. 
Earlier works on random Ising models \cite{McCoy,Griffiths} 
deserve to be mentioned.  

The two-dimensional (2D) Ising model is a marginal case. 
The specific heat diverges logarithmically at the critical point, 
that is, $\alpha_{\rm pure}=0$.  We cannot tell whether the randomness 
is relevant or not from the Harris criterion. 
Using the quantum-field theory, Dotsenko and Dotsenko \cite{DoDo} 
calculated the correlation length of the 2D diluted Ising system. 
They showed that the randomness is irrelevant but there appear 
logarithmic corrections. 
The expression for the correlation length in the weak-dilution 
region is given by
\begin{equation}
 \xi \propto \frac{(1+\lambda\ln(1/|t|))^{1/2}}{|t|},
     \quad t=(T-T_c)/T_c,
\label{eq_DD}
\end{equation}
where $\lambda (\ll 1)$ is the strength of randomness. 
Shalaev, Shankar, and Ludwig \cite{Shalaev,Shankar,Ludwig} 
proceeded with the calculation of magnetization, 
\begin{equation}
 m \propto \frac{|t|^{1/8}}{(1+\lambda\ln(1/|t|))^{1/16}}, 
\label{eq_SSl}
\end{equation}
which also shows the logarithmic corrections to the pure
Ising model without any change in critical exponents. 

The crossover is an interesting subject especially for random 
spin systems.  The critical behavior is controlled by a relevant 
fixed point.  However, it is influenced by irrelevant fixed points. 
Thus, as the system size becomes large, we observe the crossover 
behavior at the critical region between the relevant fixed point 
and irrelevant fixed points.  
In the case of diluted spin models, the magnetic order disappears 
at the percolation threshold even at $T=0$.  Then, the crossover 
between the pure fixed point, the random fixed point (if there exists), 
and the percolation fixed point is the subject of concern \cite{Cardy}.  

The problem of self-averaging is also of current interest 
for random spin systems \cite{AhHa,WiDo} because each sample 
has a different random configuration. 
The system is said to exhibit self-averaging if the relative 
variance of the thermal average of a quantity goes to zero
as the system size becomes infinite. 
Using the RG, Aharony and Harris (AH) \cite{AhHa} discussed 
the $L$ dependence of the relative cumulants of 
singular quantities, where $L$ is the linear system size; 
they discussed the self-averaging property of the random system, 
which depends on whether the randomness is relevant or not.

The Monte Carlo simulation is a powerful tool to 
study difficult problems such as random spin systems. 
However, the simulation method sometimes suffers from the problems of 
slow dynamics.  Several new algorithms for the Monte Carlo simulation 
have been proposed to overcome such difficulties. 
Cluster algorithms \cite{SwWa,Wolff} are examples of such efforts. 
The histogram method \cite{FeSw} enables us to calculate physical 
quantities for different parameters with a single simulation 
by using the reweighting technique. 

Quite recently the present authors have proposed a new effective 
cluster algorithm, which is called the probability-changing cluster 
(PCC) algorithm \cite{PCC}, of tuning the critical point automatically. 
The invaded cluster algorithm \cite{IC} was also proposed 
to determine the critical point automatically, but its ensemble 
is not necessarily clear.  In contrast, with the PCC algorithm
we approach the canonical ensemble asymptotically; we can use 
the finite-size scaling (FSS) analysis for physical quantities 
near the critical point. 
The PCC algorithm is quite useful for studying the random spin 
systems, where the distribution of the critical temperature, $T_c$, 
due to the randomness is important, because we can tune 
the critical point of each random sample automatically. 
Wiseman and Domany \cite{WiDo} pointed out the importance of 
determining the sample-dependent $T_c(L)$ in the simulational 
study of random spin systems.   They applied a reweighting technique
to search critical points of random samples, but the iteration 
process was needed and there was a difficulty in tuning $T_c(L)$ 
for some exceptional samples. 
The importance of the random average {\it after} finding 
the critical point of each sample was also pointed out 
by Bernardet {\it et al.} \cite{Bernardet}. 

In this paper, we study the 2D site-diluted Ising model 
using the PCC algorithm.  
We focus on the crossover phenomena and the sample dependence of
physical quantities. The rest of the paper is organized as follows. 
In Sec. II, we describe the model and the simulation method, the PCC 
algorithm.  In Sec. III, we address the FSS analysis of the data 
and the crossover between fixed points, paying attention to 
the corrections to FSS.  
In Sec. IV, we study the variance of the quantities, and discuss the 
self-averaging property of the 2D site-diluted Ising model.  
The summary and discussion are given in Sec. V. 

\section{Model and Simulation method}

We are concerned with the site-diluted Ising model whose Hamiltonian 
is given by
\begin{equation}
 {\cal H} = -J\sum_{\left<i,j\right>}\epsilon_i \epsilon_j S_i S_j.
\end{equation}
Here $S_i$ is the Ising spin on the lattice site $i$, and $\epsilon_i$ 
is a random variable that takes 1 (spin) or 0 (vacancy).  The summation 
is taken over the nearest-neighbor pairs $\left<i,j\right>$. 
The concentration of the spin will be denoted by $p$. 
The 2D site-diluted Ising model was already studied by 
the Monte Carlo methods \cite{Balles}, and the 2D random-bond 
Ising model was investigated by the Monte Carlo simulations 
\cite{Bernardet,WSDA,SST,Talapov,Kim94,Kim00}, the transfer-matrix 
calculation \cite{Reis}, and the high-temperature expansion \cite{Roder}. 
In the present study we give special attention to the distribution 
of physical quantities using the PCC algorithm.  

Here we briefly describe the idea of the PCC algorithm \cite{PCC}. 
The PCC algorithm is an extended version of the Swendsen-Wang 
cluster algorithm \cite{SwWa}, where the Kasteleyn-Fortuin (KF) 
representation \cite{KF} of the Ising (Potts) model is used.  
To form a cluster, parallel spins are connected with 
the probability $p^{\rm KF}=1-e^{-2J/k_BT}$. 
For the diluted model, of course, only pairs of spin sites 
are connected.  In the PCC algorithm we change the probability 
$p^{\rm KF}$ depending on the observation whether clusters are 
percolating or not percolating. 
A simple negative feedback mechanism together with the 
FSS property of the percolation leads to the determination 
of the critical point.  As $\Delta p^{\rm KF}$, 
the amount of the change of $p^{\rm KF}$, becomes small, 
the distribution of $p^{\rm KF}$ becomes a sharp Gaussian 
distribution around the mean value $p^{\rm KF}_c(L)$, 
which results in the determination of the critical 
temperature $T_c(L)$.  We approach the canonical ensemble 
in this limit.  For the more detailed description of the PCC algorithm, 
see Ref.~\cite{PCC}. 

We have made simulations for the 2D site-diluted Ising model 
on the square lattice with the system sizes 
$L=16$, 32, 64, 128 and 256.  For the spin concentration $p$, 
we have picked up $p=0.7$ and $p=0.9$. 
We deal with the grand-canonical ensemble of samples 
in which the occupation of each site is determined 
independently with probability $p$. 
The percolation threshold for the site percolation 
of the square lattice is known to be 0.592746 \cite{Ziff}.  
The final value of $\Delta p^{\rm KF}$ 
has been chosen as $\Delta p^{\rm KF} = a/L$, where $a = 6.25 \times 10^{-5}$ 
for the system size $L$. 
There are several choices for the criterion to determine percolating.  
We have employed the topological rule in the present study. 
The topological rule is that some cluster winds around the system 
in at least one of the $D$ directions in $D$-dimensional systems. 
The random average is taken over 2,000 samples for $L=16$ to 128, 
and 1,000 samples for $L=256$. 
For each random sample we have made 10,000 Monte Carlo sweeps 
to take the thermal average.

\section{Finite-Size Scaling Analysis and Crossover}

\begin{figure}
\epsfxsize=0.95\linewidth
\centerline{\epsfbox{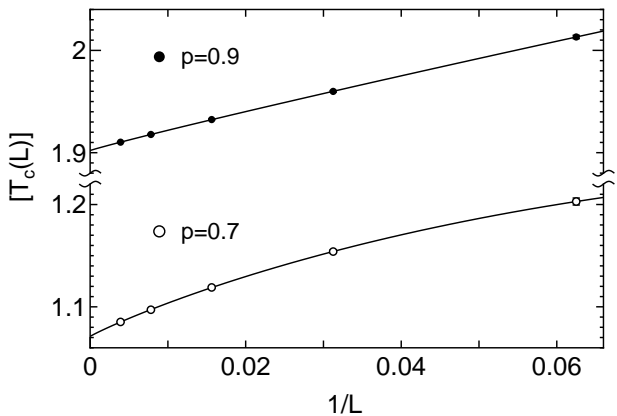}}
\vspace{2mm}
\caption{
Size dependence of critical temperatures $[T_c(L)]$ 
(in units of $J/k_B$) 
for $p=0.7$ and 0.9. 
The system sizes $L$ are 16, 32, 64, 128, and 256. 
The solid curves are the best-fitted curves using 
Eqs.~(\ref{eq_FSS_T1}) and (\ref{eq_FSS_T2}) 
for $p=0.7$ and 0.9, respectively.
}
\label{fig_1}
\end{figure}

Let us start with the size dependence of the critical temperature. 
We plot $[T_c(L)]$ as a function of $1/L$ for both $p=0.7$ and 
$p=0.9$ in Fig.~\ref{fig_1}, where the brackets $[\cdots]$ 
represent the random sample average. 
From now on, we represent the temperature in units of $J/k_B$.
The error bars are within the size of marks. 

We employ the FSS analysis for $[T_c(L)]$. 
According to the theory of FSS \cite{Fisher},  
if a quantity $Q$ has a singularity of the form 
$Q(t) \sim t^{x}$ near the criticality $t=0$, 
the corresponding quantity $Q(L,t)$ for 
the finite system with the linear size $L$ has a scaling form 
\begin{equation}
 Q(L,t) \sim L^{-x/\nu} f(tL^{1/\nu}),
\label{eq_FSS}
\end{equation}
where $\nu$ is the correlation-length exponent. 
Then, if the corrections are negligible, the critical temperature 
$T_c$ for the infinite system can be estimated 
through the relation
\begin{equation}
 \left[T_c(L)\right] = T_c + A L^{-1/\nu}.
\label{eq_FSS_T}
\end{equation}
But the corrections to FSS due to irrelevant fixed points are 
important especially in the strong-dilution region. 
Then, we use the relation 
\begin{equation}
 \left[T_c(L)\right] = T_c + A_1L^{-1/\nu}(1+B_1L^{-\Omega})
\label{eq_FSS_T1}
\end{equation}
to determine $T_c$ and $\nu$ for $p=0.7$. 
Using the non-linear fitting, we estimate $T_c$ as 1.0712(5) 
and $1/\nu$ as 0.96(1) for $p=0.7$.  Here, the number 
in the parentheses denotes the uncertainty in the last digits. 
The correction-to-scaling exponent $\Omega$ is estimated as 0.63(20). 
In the weak-dilution region ($p=0.9$), the corrections to FSS 
given in Eq.~(\ref{eq_FSS_T1}) are rather small.  However, 
there appear small logarithmic corrections, 
as was pointed out by Dotsenko and Dotsenko \cite{DoDo}, 
around the pure Ising fixed point; 
for finite systems the corrections to FSS due to 
the logarithmic term in Eq.~(\ref{eq_DD}) may be ascribed to 
the corrections to the critical exponent.  
Thus, we use the equation 
\begin{equation}
 \left[T_c(L)\right] = T_c + a_1L^{-1/\nu+b_1(\ln L)^{-\omega}}
\label{eq_FSS_T2}
\end{equation}
for the analysis of the data for $p=0.9$.  
A similar analysis was employed by Talapov {\it et al.}~\cite{Talapov} 
with $\omega=1$.  Here, we regard $\omega$ as a free parameter 
to include higher-order contributions.  
Our estimate of $T_c$ with the non-linear fitting is 1.9022(6) 
and that of $1/\nu$ is 1.00(1) for $p=0.9$.  
The solid curves in Fig.~\ref{fig_1} are 
the best-fitted curves using Eq.~(\ref{eq_FSS_T1}) 
and Eq.~(\ref{eq_FSS_T2}) for $p=0.7$ and $p=0.9$, respectively. 
Since the critical exponent $1/\nu$ for the 2D Ising model is 1, 
our estimates of $1/\nu$ for the 2D site-diluted Ising model show 
that the critical phenomena are controlled by the pure Ising 
fixed point.  

\begin{figure}
\epsfxsize=0.95\linewidth
\centerline{\epsfbox{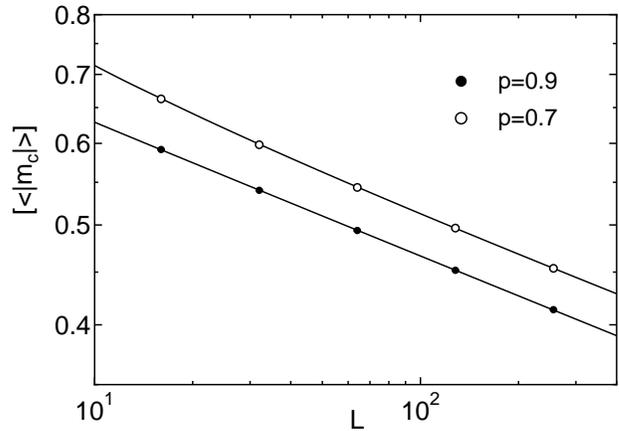}}
\vspace{2mm}
\caption{Logarithmic plot of $[\left< |m_c(L)| \right>]$ 
versus $L$ for $p=0.7$ and 0.9. 
The system sizes $L$ are 16, 32, 64, 128, and 256. 
The solid curves are computed by the non-linear fitting using 
Eqs.~(\ref{eq_FSS_M1}) and (\ref{eq_FSS_M2}) 
for $p=0.7$ and 0.9, respectively.
} 
\label{fig_2}
\end{figure}

Next turn to the average value of magnetization 
at the critical temperature, $[\left< |m_c(L)| \right>]$.  
Here, the angular brackets $\left< \cdots \right>$ 
represent the thermal average. 
We have measured the magnetization at the sample-dependent 
critical temperature $T_c(L)$ for each sample; then 
we have taken the sample average. 
The efficiency of this procedure for high-precision analysis 
was pointed out by Wiseman and Domany \cite{WiDo}. 
The log-log plot of 
$[\left< |m_c(L)| \right>]$ versus $L$ for $p=0.7$ and 0.9 
is given in Fig.~\ref{fig_2}.  The error bars are again 
within the size of marks. 

The FSS relation including the corrections to scaling 
for the magnetization may be described as 
\begin{equation}
 [\left< |m_c(L)| \right>] = A_2 L^{-\beta/\nu}(1+B_2L^{-\Omega})
 \label{eq_FSS_M1}
\end{equation}
for the strong-dilution region.  We can estimate the exponent 
$\beta/\nu$ by using Eq.~(\ref{eq_FSS_M1}). 
We should note that we do {\it not} need the information of $T_c$ 
for the estimate of $\beta/\nu$. 
We estimate the critical exponent $\beta/\nu$ as
0.123(2) for $p=0.7$ using the non-linear fitting. 
The estimated value of $\beta/\nu$ is consistent with 
that of the pure Ising model, 1/8=0.125. 
The estimate of the corrections-to-scaling exponent $\Omega$ is 
0.66(4), which is the same as that for the case of $[T_c(L)]$. 
In the weak-dilution region, we take account of the logarithmic 
corrections \cite{Shalaev,Shankar,Ludwig}.  For finite systems, 
we use the fitting function 
\begin{equation}
 [\left< |m_c(L)| \right>] = a_2 L^{-\beta/\nu+b_2(\ln L)^{-\omega}},
\label{eq_FSS_M2}
\end{equation}
which is similar to Eq.~(\ref{eq_FSS_T2}). 
The non-linear fitting based on Eq.~(\ref{eq_FSS_M2}) have led 
to the estimate of $\beta/\nu$ as 0.125(1) for $p=0.9$. 

For both cases of $[T_c(L)]$ and $[\left< |m_c(L)| \right>]$, 
the FSS analysis taking account of the corrections to scaling 
has yielded the critical exponents of the pure Ising model, 
which shows that the critical phenomena of the 2D site-diluted 
Ising model are controlled by the pure fixed point. 
In the strong-dilution region, the corrections due to 
the irrelevant fixed point are important, whereas 
in the weak-dilution region, the logarithmic corrections 
around the pure Ising fixed point are dominant. 

The effects of irrelevant fixed points also appear 
as the crossover phenomena. 
In the present case, the percolation fixed point is 
such a fixed point.  In order to see the crossover explicitly, 
we study the size dependence of the Binder parameter \cite{Binder}
defined as
\begin{equation}
 g = \frac{1}{2} 
 \left( 3 - \frac{\left< m^4 \right>}{\left< m^2 \right>^2} \right), 
\end{equation}
which is often used in the analysis of the FSS. 
Since the prefactors of the $L$ dependence in Eq.~(\ref{eq_FSS}) 
are canceled out, one may determine the critical point 
from the crossing point of the data of temperature dependence 
for different sizes as far as the correction to FSS are negligible. 
\begin{figure}
\epsfxsize=0.95\linewidth
\centerline{\epsfbox{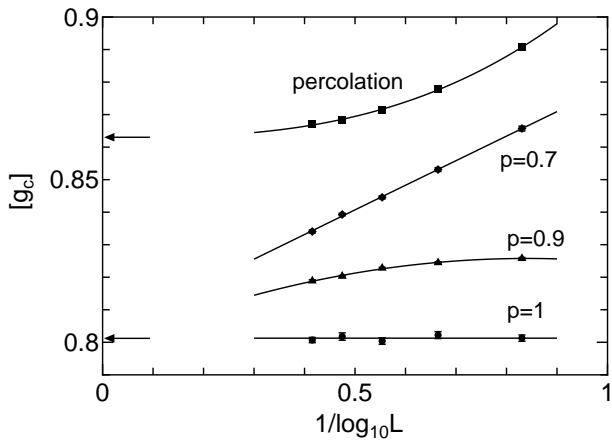}}
\vspace{2mm}
\caption{Sample average of the Binder parameter $[g_c]$ 
for $p=0.7$, 0.9 and 1.  The Binder parameter for 
the percolation problem is also shown. 
Solid curves are guiding the eye.
}
\label{fig_3}
\end{figure}

Here, we investigate the sample average of 
the Binder parameter at each $T_c(L)$, $[g_c]$.
We plot $[g_c]$ for $p=0.7$, 0.9 and 1 in Fig.~\ref{fig_3}, 
where $p=1$ is nothing but the pure Ising model. 
The error bars are within the size of marks. 
As for the size dependence, the $\log L$ contributions may appear, 
although the explicit form of $L$ dependence is not clear; 
we plot $[g_c]$ as a function of $1/\log_{10}L$ in Fig.~\ref{fig_3}, 
and the solid curves are guiding the eye. 
We also show $[g_c]$ for the percolation problem at the 
sample-dependent percolation threshold, $p_c(L)$, in Fig.~\ref{fig_3}. 
We define the magnetization for the geometric percolation problem 
by assigning the Ising spins on each cluster randomly with the 
probability of 1/2.  
The values of $[g_c]$ extrapolated to $L=\infty$ for $p=1$ and 
percolation are shown by the arrows.
We see from Fig.~\ref{fig_3} that the Binder parameter of the 
dilute Ising model approaches the value of the pure Ising model 
as the system size becomes large. 
It is the manifestation of the crossover from the percolation 
fixed point to the pure Ising fixed point with the system size. 

The fraction of lattice sites in the largest cluster, $c$, 
plays a role of order parameter in the percolation problem.  
The moment ratio of $c$, $\left<c\right>^2/\left<c^2\right>$, 
has the same FSS property as the Binder parameter because 
the prefactors of the $L$ dependence are canceled out \cite{TOH}.  

\begin{figure}
\epsfxsize=0.95\linewidth
\centerline{\epsfbox{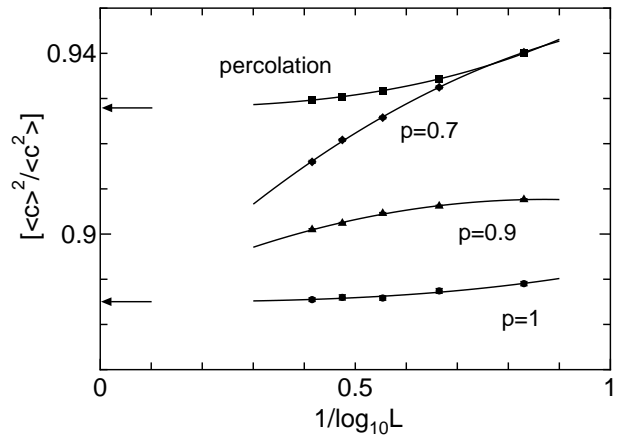}}
\vspace{2mm}
\caption{Sample average of the moment ratio for 
the fraction of lattice sites in the largest cluster 
$[\left<c\right>^2/\left<c^2\right>]$ 
for $p=0.7$, 0.9 and 1.  The data for 
the percolation problem are also shown. 
Solid curves are guiding the eye.
}
\label{fig_4}
\end{figure}

We plot the sample average $[\left<c\right>^2/\left<c^2\right>]$ 
at the critical temperature of each sample for $p=0.7$, 0.9 and 1 
in Fig.~\ref{fig_4}.  For each sample, we have measured $c$ 
at each $T_c(L)$, and have taken average over random samples.  
We also give the data for the percolation problem.  
The error bars are within the size of marks. 
We show the values extrapolated 
to $L=\infty$ for $p=1$ and percolation by the arrows.
We find that the value of $[\left<c\right>^2/\left<c^2\right>]$ 
of the dilute Ising model approaches the value of the pure Ising model 
as the system size becomes large, which is the same as the case of 
the Binder parameter, $[g_c]$. 

\section{Self-averaging}

In this section, we study the distribution of physical quantities.  
We first deal with the variance of the critical temperature 
\begin{equation}
 (\Delta T_c(L))^2 = [ T_c(L)^2 ] - [ T_c(L) ]^2.
\label{eq_var}
\end{equation}
Using the RG argument, 
AH \cite{AhHa} discussed the power-law size 
dependence, that is, $(\Delta T_c(L))^2 \propto L^{-n}$ 
for large $L$. 
According to AH, the exponent $n$ depends on whether the random 
system is controlled by the pure fixed point or the random 
fixed point, that is, 
\begin{equation}
 n =\left\{
 \begin{array}{ll}
  D & \mbox{(for pure fixed point)} \\
  2/\nu_{\rm random}  & \mbox{(for random fixed point)} 
  \end{array}
 \right.
\end{equation}
where $D$ is the spatial dimension and $\nu_{\rm random}$ is 
the correlation-length exponent for the random fixed point. 

\begin{figure}
\epsfxsize=0.95\linewidth
\centerline{\epsfbox{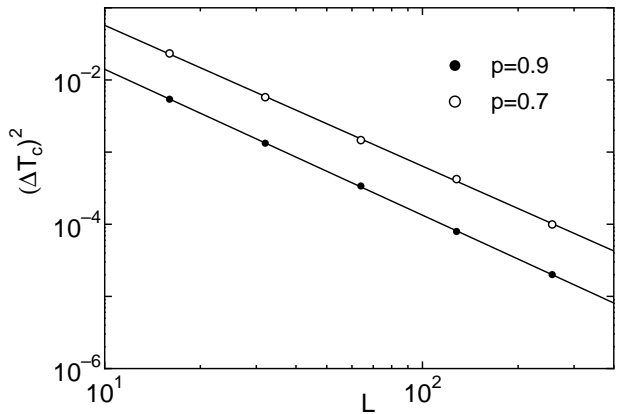}}
\vspace{2mm}
\caption{Size dependence of variance of critical temperature 
$(\Delta T_c(L))^2$, given by Eq.~(\ref{eq_var}). 
The slopes of solid lines give the exponent $n$ for $L^{-n}$; 
$n=1.95(3)$ and 2.02(3) for $p=0.7$ and 0.9, respectively. }
\label{fig_5}
\end{figure}

We plot $(\Delta T_c(L))^2$ for $p=0.7$ and 0.9 in Fig.~\ref{fig_5}. 
The variance $(\Delta T_c(L))^2$ becomes smaller 
as the system size becomes larger.  
Since the logarithmic scales are used in Fig.~\ref{fig_5}, 
we find the power-law size dependence from the linearity of 
the data; we can estimate the exponent $n$ from the slopes of lines. 
Our estimates of $n$ using least squares method are 
$n=1.95(3)$ for $p=0.7$ and $n=2.02(3)$ for $p=0.9$, respectively. 
These values are consistent with the prediction of AH \cite{AhHa}, 
that is, $n=D=2$. 
\vspace*{5mm}

We treat the relative variance for the thermal average of 
physical quantity $\left<X\right>$, 
\begin{equation}
 R_X(L) = \frac{ [ \left< X \right>^2 ] - [ \left< X \right> ]^2}
 {[ \left< X \right> ]^2},
\end{equation}
when we discuss self-averaging.  
The system is said to exhibit self-averaging if $R_X(L) \to 0$ as
$L \to \infty$.  
AH \cite{AhHa} predicted that 
the size dependence of $R_X(L)$ for the random system depends on 
whether the system is controlled by the random fixed point or 
the pure fixed point; that is, 
\begin{equation}
 R_X(L) \propto \left\{
 \begin{array}{ll}
  L^{(\alpha/\nu)_{\rm pure}} & \mbox{(for pure fixed point)} \\
  {\rm const}. \ (\ne 0)  & \mbox{(for random fixed point)} 
  \end{array}
 \right.
\end{equation}
for $L \to \infty$. 
In the case of the random fixed point, the random system 
has no self-averaging.  On the other hand, the system 
exhibits weak self-averaging in the case of the pure fixed point. 
Since the 2D Ising model is a marginal case, it is interesting to 
study the $L$ dependence of $R_X(L)$.  

\begin{figure}
\epsfxsize=0.95\linewidth
\centerline{\epsfbox{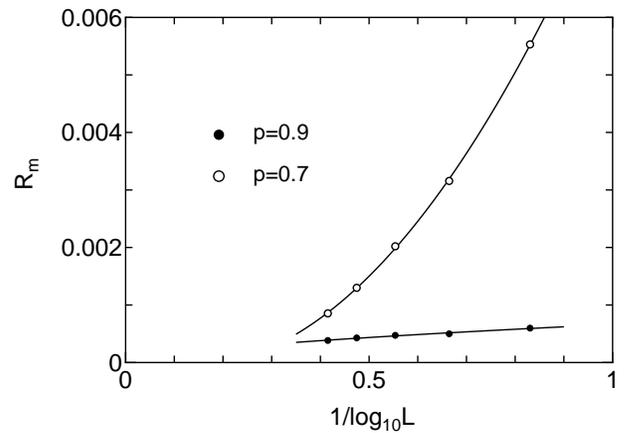}}
\vspace{2mm}
\caption{Relative variance of critical magnetization 
at the sample-dependent $T_c(L)$, $R_m$. 
Solid curves are guiding the eye. 
}
\label{fig_6}
\end{figure}

We plot $R_X(L)$ for the magnetization $m$ at the sample-dependent 
$T_c(L)$, $R_m$, as a function of $L$ in Fig.~\ref{fig_6}. 
The absolute value of $R_m$ is small for $p=0.9$.  
It has larger value for smaller $L$ for $p=0.7$, but it becomes 
smaller for larger $L$.  Since there may appear the $\log L$ 
contributions, we have chosen $1/\log_{10}L$ for the horizontal axis 
as in Figs.~\ref{fig_3} and \ref{fig_4}, but the explicit form 
of size dependence is not clear.  Anyway, from Fig.~\ref{fig_6} 
we realize $R_m\to 0$ 
as $L\to \infty$; the 2D diluted Ising model exhibits weak self-averaging.  
In other words, the thermal average of magnetization does not 
depend on samples as the system size becomes infinite. 

\section{Summary and Discussions}

To summarize, we have investigated the 2D site-diluted Ising model 
using the PCC algorithm.  Since we can tune the critical point 
of each random sample automatically, we have succeeded in 
studying the sample-dependent $T_c(L)$ and the sample average 
of physical quantities at each $T_c(L)$ systematically. 

We have used the FSS analysis for $T_c(L)$.  We have shown the 
importance of corrections to FSS due to irrelevant fixed points 
in the strong-dilution region ($p=0.7$); on the other hand, 
the logarithmic corrections are relevant 
in the weak-dilution region ($p=0.9$).  
The critical phenomena of the 2D site-diluted Ising model are 
shown to be controlled by the pure fixed point with the logarithmic 
corrections.  The crossover from the percolation fixed point 
to the pure Ising fixed point with the system size has been 
explicitly demonstrated by the study of the Binder parameter; 
it reflects the flow of renormalization. 

For the weak-dilution region, we have shown that the logarithmic 
corrections are important, which are compatible with 
the previous studies \cite{Bernardet,Balles,WSDA,SST,Talapov,Reis,Roder}.
We have also shown the crossover to the pure fixed point. 
It is opposed to the weak-universality scenario, 
that is, the exponent $\nu$ is concentration-dependent 
\cite{Kim94,Kuhn}.  No crossover was found in a recent work 
by Kim \cite{Kim00}. 
Here, we make a comment on the model.  In the present study, 
we have studied the diluted Ising model.  
Ballesteros {\it et al.}~\cite{Balles} studied the same model. 
The ground states are partial ferromagnetic states for dilution, 
and the magnetic order continuously disappears 
at the percolation threshold. 
The random bond mixture model of $J_1$ and $J_2$ couplings  
with 50 \% concentration is also used 
\cite{Bernardet,WSDA,SST,Talapov,Kim94,Kim00}
because the critical temperatures are exactly known 
by the duality relation.  The ground states of the latter model 
are always the complete ferromagnetic state except for $J_2/J_1=0$. 
In the weak randomness region, the corrections 
due to the randomness of the two models seem to be the same; 
but the crossover behavior from the percolation point may be different.
The careful study of the crossover for both models is still desired. 
Moreover, the replica symmetry breaking effects
on the critical behavior of weakly disordered systems 
were also discussed \cite{Do95a,Do95b,Do99} 
in relation to the Griffiths singularity \cite{Griffiths}.  
The check of this problem will be left to a future study.

We have studied the distribution of critical temperature $T_c(L)$. 
Its variance shows the power-law $L$ dependence, $L^{-n}$, 
and the estimate of the exponent, $n\sim 2$, is consistent 
with the prediction of AH, $n=D$.
We have also calculated the relative variance of critical 
magnetization at the sample-dependent $T_c(L)$, $R_m$.  
It becomes asymptotically close to zero 
as $L$ becomes larger even for the case of $p=0.7$. 
Thus, the 2D site-diluted Ising model is controlled by the 
pure Ising fixed point and exhibits weak self-averaging. 

In this paper we have studied the 2D diluted Ising model, 
where the pure Ising (relevant) fixed point and the percolation 
(irrelevant) fixed point are considered.  For the three-dimensional 
(3D) diluted Ising model \cite{WiDo,Heuer,Balle98,Hukushima}, 
the randomness becomes relevant because $\alpha_{\rm pure} > 0$; 
then there are three fixed points to be considered, that is, 
the random fixed point, the pure Ising fixed point, and 
the percolation fixed point.  The systematic study of the 3D diluted 
Ising model using the PCC algorithm is quite interesting, which is 
now in progress.

\section*{Acknowledgments}

We thank N. Kawashima, H. Otsuka, M. Kikuchi, and C.-K. Hu 
for valuable discussions.  
Thanks are also due to J.-S. Wang and W. Janke 
for fruitful discussions. 
The computation in this work has been done using the facilities of
the Supercomputer Center, Institute for Solid State Physics,
University of Tokyo.
This work was supported by a Grant-in-Aid for Scientific Research 
from the Ministry of Education, Science, Sports and Culture, Japan.

\end{multicols}

\vspace{-4mm}

\begin{references}
\vspace{-15mm}
\bibitem[*]{tomita} Electronic address: ytomita@phys.metro-u.ac.jp
\bibitem[\dagger]{okabe} Electronic address: okabe@phys.metro-u.ac.jp

\bibitem{Stinchcombe}
 R. B. Stinchcombe, in {\it Phase Transitions and Critical Phenomena}, 
 edited by C. Domb and J. L. Lebowitz (Academic, New York, 1983), Vol. 7; 
 in {\it Spin Glasses and Random Fields}, 
 edited by A. P. Young (World Scientific, Singapore, 1997).
 
\bibitem{Harris}
 A. B. Harris, J. Phys. C{\bf 7}, 1671 (1974).
\bibitem{McCoy}
 B. M. McCoy and T. T. Wu, Phys. Rev. {\bf 176}, 631 (1968).
\bibitem{Griffiths}
 R. B. Griffiths, Phys. Rev. Lett. {\bf 23}, 17 (1969).

\bibitem{DoDo}
 Vik. S. Dotsenko and Vl. S. Dotsenko, Sov. Phys. JETP Lett. 
 {\bf 33}, 37 (1981).
\bibitem{Shalaev}
 B. N. Shalaev, Sov. Phys. Solid State {\bf 26}, 1811 (1984);
 Physics Reports {\bf 237}, 129 (1994).
\bibitem{Shankar}
 R. Shankar, Phys. Rev. Lett. {\bf 58}, 2466 (1987); {\bf 61}, 2390 (1988).
\bibitem{Ludwig}
 A. W. W. Ludwig, Phys. Rev. Lett. {\bf 61}, 2388 (1988); 
 Nucl. Phys. B{\bf 330}, 639 (1990).
\bibitem{Cardy}
 J. Cardy, {\it Scaling and Renormalization in Statistical Physics}, 
 (Cambridge University Press, Cambridge, 1996).

\bibitem{AhHa}
 A. Aharony and A. B. Harris, Phys. Rev. Lett. {\bf 77}, 3700 (1996).
\bibitem{WiDo}
 S. Wiseman and E. Domany, Phys. Rev. Lett. {\bf 81}, 22 (1998); 
 Phys. Rev. E {\bf 58}, 2938 (1998). 

\bibitem{SwWa}
 R. H. Swendsen and J. S. Wang, Phys. Rev. Lett. {\bf 58}, 86 (1987). 
\bibitem{Wolff}
 U. Wolff, Phys. Rev. Lett. {\bf 60}, 1461 (1988). 
\bibitem{FeSw}
 A. M. Ferrenberg and R. H. Swendsen, Phys. Rev. Lett. {\bf 61}, 2635 (1988); 
 Phys. Rev. Lett. {\bf 63}, 1658(E) (1989).
\bibitem{PCC}
 Y. Tomita and Y. Okabe, Phys. Rev. Lett. {\bf 86}, 572 (2001).
\bibitem{IC}
 J. Machta, Y. S. Choi, A. Lucke, T. Schweizer, and L. V. Chayes, 
 Phys. Rev. Lett. {\bf 75}, 2792 (1995); Phys. Rev. E {\bf 54}, 1332 (1996). 

\bibitem{Bernardet}
  K. Bernardet, F. P\'azm\'andi, and G. G. Batrouni, 
  Phys. Rev. Lett. {\bf 84}, 4477 (2000).

\bibitem{Balles}
 H. G. Ballesteros, L. A. Fern\'andez, V. Mar\'{\i}n-Mayor, A. Mu\~noz Sudupe, 
 G. Parisi, and J. J. Ruiz-Lorenzo, J. Phys. A {\bf 30}, 8379 (1997).
\bibitem{WSDA}
 J. S. Wang, W. Selke, Vl. S. Dotsenko, and V. B. Andreichenko, 
 Physica A {\bf 164}, 221 (1990).
\bibitem{SST}
 W. Selke, L. N. Shchur, and A. L. Talapov, in {\it Annual Reviews of 
 Computational Physics}, edited by D. Stauffer 
 (World Scientific, Singapore, 1994), p. 17.
\bibitem{Talapov}
  A. L. Talapov and L. N. Shchur, J. Phys. C {\bf 6}, 8295 (1994).
\bibitem{Kim94}
  J.-K. Kim and A. Patrascioiu, Phys. Rev. Lett. {\bf 72}, 2785 (1994).
\bibitem{Kim00}
  J.-K. Kim, Phys. Rev. B {\bf 61}, 1246 (2000).
\bibitem{Reis}
  F. D. A. A. Reis, S. L. A. de Queiroz, and R. R. dos Santos,
  Phys. Rev. B {\bf 56}, 6013 (1997).
\bibitem{Roder}
  A. Roder, A. Adler, and W. Janke, Phys. Rev. Lett. {\bf 80}, 4697 (1998).

\bibitem{KF}  
 P. W. Kasteleyn and C. M. Fortuin, 
 J. Phys. Soc. Jpn. Suppl. {\bf 26}, 11 (1969);
 C. M. Fortuin and P. W. Kasteleyn, Physica {\bf 57}, 536 (1972).
\bibitem{Ziff}  
 R. M. Ziff, Phys. Rev. Lett. {\bf 69}, 2670 (1992).
\bibitem{Fisher}
 M. E. Fisher, in {\it Critical Phenomena, Proceedings of the International
 School of Physics ``Enrico Fermi''}, Course 51, Varenna, 1970, 
 edited by M. S. Green (Academic, New York, 1971), Vol. 51, p. 1; 
 in {\it Finite-size Scaling}, edited by 
 J. L. Cardy (North-Holland, New York, 1988).
\bibitem{Binder} 
  K. Binder, Z. Phys. B {\bf 43}, 119 (1981).
\bibitem{TOH} 
  Y. Tomita, Y. Okabe, and C.-K. Hu, 
  Phys. Rev. E {\bf 60}, 2716 (1999).

\bibitem{Kuhn}
  R. K\"uhn, Phys. Rev. Lett. {\bf 73}, 2268 (1994).

\bibitem{Do95a}
 Vik. S. Dotsenko, A. B. Harris, D. Sherrington, and  R. B. Stinchcombe, 
 J. Phys. A {\bf 28}, 3093 (1995).
\bibitem{Do95b}
 Vik. S. Dotsenko and D. E. Feldman, J. Phys. A {\bf 28}, 5183 (1995).
\bibitem{Do99}
 Vik. S. Dotsenko, J. Phys. A {\bf 32}, 2949 (1999).

\bibitem{Heuer} 
  H. O. Heuer, J. Phys. A {\bf 26}, L333 (1993).
\bibitem{Balle98}
  H. G. Ballesteros, L. A. Fern\'andez, V. Mart\'{\i}n-Mayor, 
  Mu\~noz Sudupe, G. Parisi, and J. J. Ruiz-Lorenzo,
  Phys. Rev. B {\bf 58}, 2740 (1998).
\bibitem{Hukushima}
 K. Hukushima, J. Phys. Soc. Jpn. {\bf 69}, 631 (2000).

\end{references}
\end{document}